\documentclass[dvipdfmx]{article}
\usepackage{color}

\begin{document}

\begin{center}

{\large 
Conserved Charges of Series of Solvable Lattice Models}
\vspace{0.6cm}

Kazuhiko Minami
\vspace{0.6cm}

Graduate School of Mathematics, Nagoya University, \\
Furo-cho, Chikusa-ku, Nagoya, Aichi, 464-8602, JAPAN.

\end{center}

\begin{abstract}
An infinite number of solvable Hamiltonians, 
including the transverse Ising chain, 
the XY chain with an external field, 
the cluster model with next-nearest-neighbor $x$-$x$ interactions, 
or with next-nearest-neighbor $z$-$z$ interactions, 
and other solvable models 
that can be mapped to the free fermion system 
are considered. 
All the conserved charges of these models 
written by the string-type products of the interactions 
are obtained. 
In the case of the transverse Ising chain, 
all the known charges are rederived, 
and in the case of the other models, 
new conserved charges are obtained. 
\end{abstract}


\noindent
e-mail: minami@math.nagoya-u.ac.jp

\section{Introduction}
\label{introduction}

The existence of locally conserved quantities 
is sometimes investigated 
with respect to the thermal equilibrium of isolated quantum systems;  
they yield the absence of thermal equilibrium. 
The existence and absence are also important with respect to chaos, 
and the validity of the linear response theory.
It is easy to construct an infinite number of operators 
that commute with a given Hamiltonian. 
In particular, when the Hamiltonian ${\cal H}$ is diagonalizable, 
then the operators which have the same eigenvectors with ${\cal H}$ 
and have arbitrary eigenvalues commute with ${\cal H}$. 
But what is important is 
the existence or non-existence of local conserved charges 
that are described by physical observables.  

In the case of the free fermion system 
\begin{eqnarray}
{\cal H}
=
\sum_{j=1}^{N}\epsilon_{j}c_{j}^{\dag}c_{j},
\label{Ham-fermion}
\end{eqnarray}
where $c_{j}$'s are the fermi operators, 
and when we consider the operators of the form 
\begin{eqnarray}
\prod_{j=1}^{N}q_{j},
\hspace{0.6cm}
{\rm where}
\hspace{0.3cm}
q_{j}=1_{j},\: c_{j, },\: c_{j}^{\dag},\: c_{j}^{\dag}c_{j},
\label{fermion-op-conserve}
\end{eqnarray}
it is easy to convince that 
the operators (\ref{fermion-op-conserve}) commute with (\ref{Ham-fermion}) 
if and only if the condition  
\begin{eqnarray}
\sum_{j=1}^{N}l_{j}\epsilon_{j}=0, 
\hspace{0.6cm}
{\rm where}
\hspace{0.3cm}
[c_{j}^{\dag}c_{j},\: q_{j}]=l_{j}q_{j}
\end{eqnarray}
is satisfied. 
However, some non-local transformation is usually required 
between the operators $c_{j}$ and the operators of current physical interest, 
for example the Pauli spin operators, 
and therefore it is not easy to convince 
whether the operators (\ref{fermion-op-conserve}) are local or not. 


Fagotti and Essler\cite{FagottiEssler13} and 
Fagotti\cite{Fagotti14} 
considered a general quadratic form in Majorana fermions 
\begin{eqnarray}
{\cal H}
=
\frac{1}{2}
\sum_{l,n}a_{l}{\cal H}_{ln}a_{n},
\label{Ham-Fagotti}
\end{eqnarray}
and derived that the general form of the conserved charges 
are given by  
\begin{eqnarray}
I_{r}
=
\frac{1}{2}
\sum_{l,n}a_{l}{\cal I}_{r;ln}a_{n},
\label{Charge-Fagotti}
\end{eqnarray}
where ${\cal I}_{r;ln}$ satisfy certain commutation relations. 
In \cite{FagottiEssler13} and \cite{Fagotti14}, 
the localness of the conserved charges was also investigated. 



As for the transverse Ising chain, 
conserved charges have been investigated and obtained explicitly. 
Grady\cite{Grady82} obtained the conserved charges 
through an equivalence between two separate transverse Ising chains 
and the XYZ chain with the coupling $J_{\rm Y}$ being zero. 
An infinite set of local conserved charges of the transverse Ising chain 
were obtained from those of the XYZ model 
through this equivalence. 
Doran and Grady\cite{DolanGrady82} 
constructed an infinite set of conserved charges 
from a self-duality and a relation 
which is now known as the Dolan-Grady condition. 
Their charges form a subset of the charges found in \cite{Grady82}.
The charges found in \cite{Grady82} and \cite{DolanGrady82} 
are linear combinations of string-type operators, 
which can already be seen for example in \cite{Suzuki71}. 
Uglov and Ivanov\cite{UglovIvanov96} 
considered an $sl(n)$ analogue of the Onsager algebra, 
which is isomorphic to a subalgebra of the $sl(n)$ loop algebra for $n\geq 3$, 
and obtained a family of conserved charges. 
The transverse Ising chain with inhomogeneities satisfies their condition. 
Prosen\cite{Prosen98} 
introduced a Hamiltonian which includes the interactions 
considered in \cite{Suzuki71}, 
and obtained two series of conserved charges. 
Fagotti and Essler\cite{FagottiEssler13} 
derived the conserved charges of the transverse Ising chain,  
and Fagotti\cite{Fagotti14} derived those of the XY chain, 
from the general framework (\ref{Charge-Fagotti}). 
Chulliparambil et al.\cite{Chulliparambil23} 
introduced an $SO(n)$ generalization of the transverse Ising chain,
and investigated its structures 
including the existence of an infinite number of conserved charges. 
Chiba\cite{Chiba24} 
considered the transverse Ising chain with a longitudinal field, 
and derived that there exist no local conserved charges under certain conditions, 
and also obtained conserved charges when there is no longitudinal field. 
Nozawa and Fukai\cite{NozawaFukai20} derived the conserved charges 
of the $s=1/2$ XXZ chain.

Anshelevich and Gusev\cite{AnshelevichGusev81} 
considered the transverse Ising chain on an infinite lattice ${\bf Z}$, 
and derived the conserved charges. 
Gusev\cite{Gusev82} (see also \cite{GrabowskiMathieu95}) 
derived the explicit expression of the conserved charges
of the XY chain on an infinite lattice ${\bf Z}$.  
Lie algebraic structures of the conserved charges of the XY model 
on an infinite lattice ${\bf Z}$ 
were investigated by Araki\cite{Araki90}. 

The absence of conserved charges has also been derived 
in many one-dimensional quantum chains, e.g.  
XYZ chain with a magnetic field\cite{Shiraishi19}, 
mixed-field Ising chain\cite{Chiba24}, 
PXP model\cite{ParkLee24}, 
$s=1/2$ XYZ chain 
with next-nearest-neighbor interactions\cite{Shiraishi24}, 
and 
$s=1/2$ XYZ chain 
with arbitrary magnetic field\cite{YamaguchiChibaShiraishi24}.


Recently, a fermionization method, 
which can be regarded 
as an algebraic generalization of the Jordan-Wigner transformation, 
was introduced in \cite{Minami16}. 
In this formula, the transformation that diagonalizes the Hamiltonian 
is automatically obtained from the Hamiltonian, 
if the Hamiltonian satisfies the condition (\ref{cond}). 
Through this formula, 
an infinite number of solvable lattice models were introduced 
in \cite{Minami17}. 
The transverse Ising chain, the Kitaev chain, the XY chain with an external field 
can be solved through this formula, 
and in these cases the transformation 
results in the ordinal Jordan-Wigner transformation. 
Other models such as the cluster model 
with the next-nearest-neighbor $x$-$x$ interactions, 
or with the next-nearest-neighbor $z$-$z$ interactions, 
the Jordan-Wigner transformation does not work, 
but new transformations can be obtained 
and the models are diagonalized. 
This formula was applied to 
one-dimensional quantum spin chains\cite{Minami17}\cite{YanagiharaMinami20}, 
to the honeycomb-lattice Kitaev model 
with the Wen-Toric-code interactions\cite{Minami19}.  
A graph-theoretic formulation was also introduced\cite{OguraImamuraKameyama20}, 
and the relation with the Onsager algebra was considered\cite{Minami21}. 
%

Formula in \cite{FagottiEssler13} and \cite{Fagotti14} 
for the free fermion system are general, 
but it is not always easy for practical Hamiltonians 
to derive ${\cal H}_{ln}$ and write down the conserved charges 
explicitly in a simple form. 
In this short note,
the conserved charges of the Hamiltonian 
that are given in a form (\ref{hamKjetaj}) and (\ref{cond}) 
are considered, 
and all the conserved charges composed of string-type operators 
are determined. 
In the case of the transverse Ising chain and the XY chain, 
all the known conserved charges are re-derived, 
and in the case of the other models, 
new conserved charges are obtained. 


\section{Hamiltonian and conserved charges}

\subsection{Fermionization method and Hamiltonian}

Let us consider the Hamiltonian 
\begin{eqnarray}
-\beta{\cal H}
=
\sum_{j=1}^{M}K_j \eta_j,
\label{hamKjetaj}
\end{eqnarray}
where the periodic boundary condition is assumed, 
and the operators $\{\eta_j\}$ $(j=1, 2, \ldots, M)$ satisfy 
the relation 
\begin{eqnarray}
\eta_{j}\eta_{k}
=
\left\{
\begin{array}{cl}
1 & j=k \\
-\eta_{k}\eta_{j} & \eta_{j}\: {\rm and}\: \eta_{k} \:{\rm are}\: {\rm adjacent} \\
\eta_{k}\eta_{j} &{\rm otherwise.} \\
\end{array}
\right.
\label{cond}
\end{eqnarray}
The operators $\eta_j$ and $\eta_k$ are called 'adjacent' 
when $(j, k)=(j, j+1)$ $(1\leq j\leq M-1)$, or $(j, k)=(M, 1)$. 
It was shown in \cite{Minami16} 
that this Hamiltonian can be diagonalized, 
and the free energy can be obtained 
using only the relation (\ref{cond}). 
Operators $\{\eta_j\}$ are arbitrary 
provided they satisfy (\ref{cond}).
The Hamiltonian (\ref{hamKjetaj}) is mapped 
to a Hamiltonian of the free fermion system by the transformation 
\begin{eqnarray}
\varphi_j
=
\frac{1}{\sqrt{2}}
e^{i\frac{\pi}{2}(j-1)}
\eta_0
\eta_1
\eta_2
\cdots
\eta_j
\hspace{0.8cm}
(0\leq j\leq M-1), 
\label{transmain}
\end{eqnarray}
where $\eta_0$ is an initial operator satisfying 
$\eta_0^2=-1$,  $\eta_0\eta_1=-\eta_1\eta_0$,  
and $\eta_0\eta_k=\eta_k\eta_0$ $(2\leq k\leq M)$, 
which is introduced to avoid unnecessary boundary effects. 
We find $(-2i)\varphi_{j}\varphi_{j+1}=\eta_{j+1}$ 
and 
\begin{eqnarray}
\{\varphi_j, \varphi_k\}
=
\varphi_j\varphi_k+\varphi_k\varphi_j
=
\delta_{jk}
\label{anticom}
\end{eqnarray}
for all $j$, $k$. 
Then we find the Hamiltonian (\ref{hamKjetaj}) is written 
as a sum of two-body products of $\varphi_{j}$, 
which are the Majorana fermion operators, 
and thus can be diagonalized. 
The transformation (\ref{transmain}) is automatically obtained 
from the original Hamiltonian. 

When we consider the operators 
$\eta_{2j-1}^{(1)}=\sigma_{j}^z$, 
$\eta_{2j}^{(1)}=\sigma_{j}^x\sigma_{j+1}^x$ for $j\geq 1$, 
$\eta_{0}^{(1)}=i\sigma_{1}^x$, 
and $M=2N$, where $N$ is the number of sites, 
they satisfy the condition (\ref{cond}).  
In this case, 
(\ref{hamKjetaj}) is the one-dimensional transverse Ising model, 
and (\ref{transmain}) results in the Jordan-Wigner transformation.

When we consider the operators 
$\eta_{2j-1}^{(2)}=\sigma_{2j-1}^x\sigma_{2j}^x$, 
$\eta_{2j}^{(2)}=\sigma_{2j}^y\sigma_{2j+1}^y$ for $j\geq 1$, 
$\eta_{0}^{(2)}=i\sigma_{1}^y$, 
and $M=N$, 
they also satisfy the condition (\ref{cond}). 
In this case, (\ref{hamKjetaj}) is the Kitaev chain, 
and we again find the Jordan-Wigner transformation from (\ref{transmain}). 
These facts immediately yield the result 
that the free energies of the transverse Ising chain and the Kitaev chain 
are identical, 
because there remains no difference 
when these two models are written by the operators $\eta_{j}$.  
In the case of the Kitaev chain, 
we can introduce another series of operators 
$\zeta_{2j-1}^{(2)}=\sigma_{2j-1}^y\sigma_{2j}^y$, 
$\zeta_{2j}^{(2)}=\sigma_{2j}^x\sigma_{2j+1}^x$ for $j\geq 1$, 
and $\zeta_{0}^{(2)}=i\sigma_{1}^x$. 
Then $\eta_{j}^{(2)}$ $(j\geq 1)$ and $\zeta_{k}^{(2)}$ $(k\geq 1)$ 
commute with each other. 
When we consider the Hamiltonian
\begin{eqnarray}
-\beta{\cal H}
=
-\beta{\cal H^{\eta}}-\beta{\cal H^{\zeta}},
\hspace{0.6cm}
-\beta{\cal H^{\eta}}
=
\sum_{j=1}^{M}K_j \eta_j,
\hspace{0.6cm}
-\beta{\cal H^{\zeta}}
=
\sum_{j=1}^{M}K_j \zeta_j,
\label{hamXY}
\end{eqnarray}
${\cal H^{\eta}}$ and ${\cal H^{\zeta}}$ are the Kitaev chains, respectively, 
$[{\cal H^{\eta}}, {\cal H^{\zeta}}]=0$, 
and are diagonalized by the transformation (\ref{transmain}) independently. 
The total Hamiltonian ${\cal H}$ forms the XY chain.

When we consider the operators 
$\eta_{2j-1}^{(3)}=\sigma_{2j-1}^x\sigma_{2j}^z\sigma_{2j+1}^x$, 
$\eta_{2j}^{(3)}=\sigma_{2j}^x 1_{2j+1}\sigma_{2j+2}^x$ for $j\geq 1$, 
$\eta_{0}^{(3)}=i\sigma_{2}^x$, 
and $M=N$, 
they satisfy the condition (\ref{cond}). 
We can introduce another series of operators 
$\zeta_{2j-1}^{(3)}=\sigma_{2j}^x\sigma_{2j+1}^z\sigma_{2j+2}^x$,  
$\zeta_{2j}^{(3)}=\sigma_{2j+1}^x 1_{2j+2}\sigma_{2j+3}^x$ for $j\geq 1$, 
and $\zeta_{0}^{(3)}=i\sigma_{3}^x$. 
The total Hamiltonian forms the cluster model 
with the next-nearest-neighbor $x$-$x$ interactions, 
which cannot be diagonalized by the Jordan-Wigner transformation, 
but diagonalized by the transformation (\ref{transmain}) as shown in \cite{YanagiharaMinami20}.

When we consider the operators 
$\eta_{2j-1,k}^{(4)}
=\sigma_{4j-4+k}^x\sigma_{4j-3+k}^z\sigma_{4j-2+k}^x$, 
$\eta_{2j, k}^{(4)}
=\sigma_{4j-2+k}^z 1_{4j-1+k}\sigma_{4j+k}^z$ for $j\geq 1$, 
$\eta_{0, k}^{(4)}=i\sigma_{k+1}^x$, where $k=1, 2, 3, 4$, 
and $M=N/2$, 
we find four Hamiltonians which commute with each other, 
and the total Hamiltonian forms the cluster model 
with the next-nearest-neighbor $z$-$z$ interactions.  


Let us consider the case $M=2N$ 
(this restriction can be removed), 
and assume that $N/2$ and $M/4$ are positive integers. 
Let us introduce the notation  
$\varphi_{2j-2}=\varphi_1(j)$ and $\varphi_{2j-1}=\varphi_2(j)$. 
In the case of $\{\eta_j^{(1)}\}$, 
we find  
$(-2i)\varphi^{(1)}_1(j)\varphi^{(1)}_2(j)=\eta_{2j-1}^{(1)}=\sigma_{j}^z$,  
$(-2i)\varphi^{(1)}_2(j)\varphi^{(1)}_1(j+1)=\eta_{2j}^{(1)}=\sigma_{j}^x\sigma_{j+1}^x$. 
Let us consider the Hamiltonian 
\begin{eqnarray}
-\beta{\cal H}
=
-\beta{\cal H}_{1}-\beta{\cal H}_{2}-\beta{\cal H}_{3},
\label{HamGen}
\end{eqnarray}
where each $-\beta{\cal H}_{j}$ is composed of 
successive products of $\eta_{j}$ as 
\begin{eqnarray}
-\beta{\cal H}_{1}
&=&
L_1 \sum_{j=1}^{N} \eta_{2j-1}
+K_1 \sum_{j=1}^{N} \eta_{2j},
\nonumber\\
-\beta{\cal H}_{2}
&=&
L_2 \sum_{j=1}^{N} \eta_{2j-1}\eta_{2j}
+K_2 \sum_{j=1}^{N} \eta_{2j}\eta_{2j+1},
\nonumber\\
-\beta{\cal H}_{3}
&=&
L_3 \sum_{j=1}^{N} \eta_{2j-1}\eta_{2j}\eta_{2j+1}
+K_3 \sum_{j=1}^{N} \eta_{2j}\eta_{2j+1}\eta_{2j+2},
\label{Ham123}
\end{eqnarray}
and the periodic boundary condition is assumed. 
In the case of the operators $\{\eta_j^{(1)}\}$, 
for example, 
we find that the successive products of $\eta_{j}^{(1)}$ give
\footnote{
Errors in signs and subscripts found in \cite{Minami17} 
are corrected in (\ref{series1}) and the Tables of this paper.
}
\begin{eqnarray}
(-2i)\:\varphi^{(1)}_2(j)\varphi^{(1)}_1(j+3)
&=&
\eta^{(1)}_{2j}\eta^{(1)}_{2j+1}\eta^{(1)}_{2j+2}\eta^{(1)}_{2j+3}\eta^{(1)}_{2j+4}
\hspace{0.3cm}
=
\sigma^x_{j}\sigma^z_{j+1}\sigma^z_{j+2}\sigma^x_{j+3}
\nonumber
\\
(+2i)\:\varphi^{(1)}_2(j)\varphi^{(1)}_1(j+2)
&=&
\eta^{(1)}_{2j}\eta^{(1)}_{2j+1}\eta^{(1)}_{2j+2}
\hspace{1.7cm}
=
(-1)\:\sigma^x_{j}\sigma^z_{j+1}\sigma^x_{j+2}
\nonumber
\\
(-2i)\:\varphi^{(1)}_2(j)\varphi^{(1)}_1(j+1)
&=&
\eta^{(1)}_{2j}
\hspace{3.1cm}
=
\sigma^x_{j}\sigma^x_{j+1}
\nonumber
\\
(+2i)\:\varphi^{(1)}_2(j)\varphi^{(1)}_1(j)
&=&
\eta^{(1)}_{2j-1}
\hspace{2.8cm}
=
\sigma^z_{j}
\nonumber
\\
(-2i)\:\varphi^{(1)}_2(j)\varphi^{(1)}_1(j-1)
&=&
\eta^{(1)}_{2j-3}\eta^{(1)}_{2j-2}\eta^{(1)}_{2j-1}
\hspace{1.4cm}
=
\sigma^y_{j-1}\sigma^y_{j}
\nonumber
\\
(+2i)\:\varphi^{(1)}_2(j)\varphi^{(1)}_1(j-2)
&=&
\eta^{(1)}_{2j-5}\eta^{(1)}_{2j-4}\eta^{(1)}_{2j-3}\eta^{(1)}_{2j-2}\eta^{(1)}_{2j-1}
=
(-1)\:\sigma^y_{j-2}\sigma^z_{j-1}\sigma^y_{j}
\label{series1}
\end{eqnarray}
and
\begin{eqnarray}
(-2i)\:\varphi^{(1)}_1(j)\varphi^{(1)}_1(j+2)
&=&
(-i)\:\eta^{(1)}_{2j-1}\eta^{(1)}_{2j}\eta^{(1)}_{2j+1}\eta^{(1)}_{2j+2}
=
(-1)\:\sigma^y_{j}\sigma^z_{j+1}\sigma^x_{j+2}
\nonumber
\\
(+2i)\:\varphi^{(1)}_1(j)\varphi^{(1)}_1(j+1)
&=&
(-i)\:\eta^{(1)}_{2j-1}\eta^{(1)}_{2j}
\hspace{1.4cm}
=
\sigma^y_{j}\sigma^x_{j+1}
\nonumber
\\
(-2i)\:\varphi^{(1)}_2(j)\varphi^{(1)}_2(j+1)
&=&
(+i)\:\eta^{(1)}_{2j}\eta^{(1)}_{2j+1}
\hspace{1.4cm}
=
\sigma^x_{j}\sigma^y_{j+1}
\nonumber
\\
(+2i)\:\varphi^{(1)}_2(j)\varphi^{(1)}_2(j+2)
&=&
(+i)\:\eta^{(1)}_{2j}\eta^{(1)}_{2j+1}\eta^{(1)}_{2j+2}\eta^{(1)}_{2j+3}
=
(-1)\:\sigma^x_{j}\sigma^z_{j+1}\sigma^y_{j+2}
\label{series2}
\end{eqnarray}
The spin products (\ref{series1}) are those introduced in \cite{Suzuki71}, 
and the interactions (\ref{series1}) and (\ref{series2}) 
are those considered in \cite{Prosen98} and \cite{Minami16}.
Similarly, every series of operators $\{\eta^{(k)}_j\}$ 
that satisfy (\ref{cond}) 
leads to an infinite number of solvable interactions. 
Some examples are shown in Table 1 and Table 2, 
which are originally given in \cite{Minami17}. 
If the couplings $L_{l}$ and $K_{l}$ for even $l$ are pure imaginary, 
the Hamiltonian (\ref{HamGen}) is hermitian. 
The Hamiltonian (\ref{HamGen}) includes the most important interactions,  
such as those of the transverse Ising model, 
the XY interactions, and the cluster interactions. 
Generalizations to include higher-body terms are straightforward.

Because the periodic boundary condition $\eta_{M+l}=\eta_{l}$ is assumed, 
we obtain from the definition of $\varphi_{j}$ that  
\begin{eqnarray}
\varphi_{2}(\frac{M}{2}+l)
&=&
\varphi_{M+2l-1}
\nonumber\\
&=&
\frac{1}{\sqrt{2}}
e^{i\frac{\pi}{2}(M+2l-2)}
\eta_0
(\eta_1
\eta_2
\cdots
\eta_M)
\eta_{M+1}
\cdots
\eta_{M+2l-1}
\nonumber\\
&=&
\varphi_{2l-1}
(\eta_1
\eta_2
\cdots
\eta_M)
=
\varphi_{2}(l)
(\eta_1
\eta_2
\cdots
\eta_M)
\hspace{0.8cm}
(l\geq 1), 
\label{transbc}
\end{eqnarray}
and similarly 
$\displaystyle
\varphi_{1}(\frac{M}{2}+l)
=
\varphi_{1}(l)
(\eta_1
\cdots
\eta_M)
$ 
$\;(l\geq 1)$. 
The product $\eta_1\cdots\eta_M$ commutes with $\eta_{j}$ $(j\geq 1)$, 
satisfies $(\eta_1\cdots\eta_M)^{2}=1$,
and hence its eigenvalues are $\pm 1$. 
Then we find in each subspace 
where $\eta_1\cdots\eta_M=1$ and $-1$ that 
\begin{eqnarray}
\varphi_\tau(\frac{M}{2}+l)
=
\left\{
\begin{array}{cl}
+\varphi_\tau(l) & \hspace{0.3cm}\eta_1\cdots\eta_M=+1\\
-\varphi_\tau(l) & \hspace{0.3cm}\eta_1\cdots\eta_M=-1,
\end{array}
\right.
\label{bcond}
\end{eqnarray}
for $\tau=1, 2$ and $l\geq 1$. 

Successive products of $\eta_{j}$ can then be written 
by two-body products of $\varphi_{\tau}(j)$, 
and the Hamiltonian (\ref{HamGen}) can be diagonalized, 
as shown in \cite{Minami16} and \cite{Minami17}, 
by a Fourier transformation   
\begin{eqnarray}
\varphi_\tau(j)
=
\frac{1}{\sqrt{N}}\sum_{0<q<\pi}(e^{iqj}c_\tau(q)+e^{-iqj}c_\tau^\dag(q)),
\label{Fourier}
\end{eqnarray}
where 
\begin{eqnarray}
\{c_r^\dag(p), c_s(q)\}
=
\delta_{pq}\delta_{rs},
\hspace{0.4cm}
\{c_r(p), c_s(q)\}=0.
\label{comcpcq}
\end{eqnarray}

\subsection{Conserved charges}

Let us consider the conserved charges $Q$ of the Hamiltonian (\ref{HamGen}), 
i.e. consider the operators that commute with (\ref{HamGen}). 
Let us introduce the inner derivatives $\delta_{j}$ and $\delta$ 
by the relations that 
\begin{eqnarray}
[-\beta{\cal H}_{j}, Q]=\delta_{j} Q
\hspace{0.6cm}
(j=1, 2, 3) 
\label{comH}
\end{eqnarray}
and 
\begin{eqnarray}
\delta=\delta_{1}+\delta_{2}+\delta_{3}.
\end{eqnarray}
We consider the eigenspace of $\delta$ 
corresponding to the eigenvalue $0$, 
i.e. the kernel of $\delta$. 
If and only if $Q$ belongs to ${\rm Ker\:} \delta$, 
then $[-\beta{\cal H}, Q]$ is equal to $0$, 
and $Q$ is an conserved charge of ${\cal H}$. 
We consider representation matrices of $\delta_{j}$ and $\delta$, 
which are denoted by the same symbols $\delta_{j}$ and $\delta$, 
respectively, 
acting on a subspace of the total Hilbert space, 
and determine all the conserved charges that belong to this subspace. 

The operators $1$ and $\eta_{1}\eta_{2}\cdots\eta_{M}$ 
commute with ${\cal H}$. 
Thus as a basis set for the representation matrix 
(a basis set for our subspace), 
let us consider the following $2(2N-1)$ number of 
string-type operators composed of $\eta_{j}$, i.e. 
\begin{eqnarray}
\sum_{j=1}^{N}\eta_{2j-1},
\hspace{0.3cm}
\sum_{j=1}^{N}\eta_{2j},
\hspace{0.3cm}
\sum_{j=1}^{N}\eta_{2j-1}\eta_{2j},
\hspace{0.3cm}
\sum_{j=1}^{N}\eta_{2j}\eta_{2j+1},
\hspace{0.3cm}
\sum_{j=1}^{N}\eta_{2j-1}\eta_{2j}\eta_{2j+1},
\hspace{0.3cm}
\ldots
\hspace{2.0cm}
\nonumber\\
\hspace{5.3cm}
\ldots
\hspace{0.3cm}
,
\hspace{0.3cm}
\sum_{j=1}^{N}\eta_{2j-1}\eta_{2j}\cdots\eta_{2j+2N-3},
\hspace{0.3cm}
\sum_{j=1}^{N}\eta_{2j}\eta_{2j+1}\cdots\eta_{2j+2N-2}.
\label{base}
\end{eqnarray}
The matrix $\delta_{1}$ is represented as 
\begin{small}
\begin{eqnarray}
2\times
\left(
\begin{array}{ccccccccccccccccc}
0 & 0 &-K_{1}&K_{1}&  &  &  &  &  &  &  &  &  &  &  \\
0 & 0 &L_{1}&-L_{1}&  &  &  &  &  &  &  &  &  &  &  \\
-K_{1}&L_{1}&0 &0 &-L_{1}&K_{1}&  &  &  &  &  &  &  &  &  \\
K_{1}&-L_{1}&0 &0 &L_{1}&-K_{1}&  &  &  &  &  &  &  &  &  \\
  &   &-L_{1}&L_{1}&0 &0 &-K_{1}&K_{1}&  &  &  &  &  &  &  \\
  &   &K_{1}&-K_{1}&0 &0 &L_{1}&-L_{1}&  &  &  &  &  &  &  \\
  &   &  &  &-K_{1}&L_{1}&0 &0 &  &  &  &  &  &  &  \\
  &   &  &  &K_{1}&-L_{1}&0 &0 &  &  &  &  &  &  &  \\
  &   &  &  &  &  &  &  &\ddots &  &  &  &  &  &  &  \\
  &   &  &  &  &  &  &  &  &0 &0 &-K_{1}&K_{1}&  &  \\
  &   &  &  &  &  &  &  &  &0 &0 &L_{1}&-L_{1}&  &  \\
  &   &  &  &  &  &  &  &  &-K_{1}&L_{1}&0 &0 &-L_{1}&K_{1}\\
  &   &  &  &  &  &  &  &  &K_{1}&-L_{1}&0 &0 &L_{1}&-K_{1}\\
  &   &  &  &  &  &  &  &  &  &  &-L_{1}&L_{1}&0 &0 \\
  &   &  &  &  &  &  &  &  &  &  &K_{1}&-K_{1}&0 &0 
\end{array}
\right), 
\label{delta1}
\end{eqnarray}
\end{small}
where the matrix is composed of the iterations of the matrices 
$\displaystyle
\left(
\begin{array}{cc}
-K_{1}&K_{1}\\
L_{1}&-L_{1}
\end{array}
\right)
$, 
$\displaystyle
\left(
\begin{array}{cc}
-L_{1}&K_{1}\\
L_{1}&-K_{1}
\end{array}
\right)
$, 
and 
$\displaystyle
\left(
\begin{array}{cc}
-K_{1}&L_{1}\\
K_{1}&-L_{1}
\end{array}
\right)
$, 
$\displaystyle
\left(
\begin{array}{cc}
-L_{1}&L_{1}\\
K_{1}&-K_{1}
\end{array}
\right)
$. 
When one considers, for example, the relation 
\begin{eqnarray}
[-\beta{\cal H}_{1}, \sum_{j=1}^{N}\eta_{2j-1}]
=
-K_{1}\sum_{j=1}^{N}\eta_{2j-1}\eta_{2j}
+K_{1}\sum_{j=1}^{N}\eta_{2j}\eta_{2j+1},
\label{commex}
\end{eqnarray}
one can find the first column of (\ref{delta1}). 
The matrix $\delta_{2}$ is represented as 
\begin{small}
\begin{eqnarray}
2\times
\left(
\begin{array}{ccccccccccccccccc}
0 &M_{2}&0 &0 &-M_{2}&0 &  &  &  &  &  &  &  &  &  &   \\
-M_{2}& 0 &0 &0 &0 &M_{2} &  &  &  &  &  &  &  &  &  &  &   \\
0 & 0 &0 &0 &0 &0 &0 &0 &  &  &  &  &  &  &  &  &   \\
0 & 0 &0 &0 &0 &0 &0 &0 &  &  &  &  &  &  &  &  &   \\
M_{2} & 0 &0 &0 &0 &0 &0 &0 &-M_{2} &0 &  &  &  &  &  &  &   \\
0 & -M_{2}&0 &0 &0 &0 &0 &0 &0 &M_{2} &  &  &  &  &  &  &   \\
  &   &0 &0 &0 &0 &0 &0 &  &  &  &  &  &  &  &  &   \\
  &   &0 &0 &0 &0 &0 &0 &  &  &  &  &  &  &  &  &   \\
  &   &  &  &M_{2} &0 &  &  &  &  &  &  &  &  &  &  &   \\
  &   &  &  &0 &-M_{2} &  &  &  &  &  &  &  &  &  &  &  \\
  &   &  &  &  &  &  &  &  &  &\ddots&  &  &  &  &  &  \\
  &   &  &  &  &  &  &  &  &  &  &  &  &  &  &-M_{2}&0  \\
  &   &  &  &  &  &  &  &  &  &  &  &  &  &  &0 &M_{2} \\
  &   &  &  &  &  &  &  &  &  &  &  &  &0 &0 &0 &0 \\
  &   &  &  &  &  &  &  &  &  &  &  &  &0 &0 &0 &0 \\
  &   &  &  &  &  &  &  &  &  &  &M_{2} &0 &0 &0 &0 &-M_{2} \\
  &   &  &  &  &  &  &  &  &  &  &0 &-M_{2}&0 &0 &M_{2}&0 
 \end{array}
\right),
\label{delta2}
\end{eqnarray}
\end{small}
where $M_{2}=L_{2}-K_{2}$, 
and the matrix is composed of iterations of 
$\displaystyle
\left(
\begin{array}{cc}
-M_{2}&0\\
0&M_{2}
\end{array}
\right)
$, 
and 
$\displaystyle
\left(
\begin{array}{cc}
M_{2}&0\\
0&-M_{2}
\end{array}
\right)
$. 
The matrix $\delta_{3}$ is represented as 
\begin{small}
\begin{eqnarray}
2\times
\left(
\begin{array}{cccccccccccccccc}
0 & 0 &L_{3}&-L_{3}&0 &0 &-K_{3}&K_{3}&  &  &  &  &  &  &  &  \\
0 & 0 &-K_{3}&K_{3}&0 &0 &L_{3}&-L_{3}&  &  &  &  &  &  &  &  \\
L_{3}&-K_{3}&0 &0 &0 &0 &0 &0 &-L_{3}&K_{3}&  &  &  &  &  &  \\
-L_{3}&K_{3}&0 &0 &0 &0 &0 &0 &L_{3}&-K_{3}&  &  &  &  &  &  \\
0 & 0 &0 &0 &0 &0 &0 &0 &  &  &  &  &  &  &  &  \\
0 & 0 &0 &0 &0 &0 &0 &0 &  &  &  &  &  &  &  &  \\
-K_{3}&L_{3}&0 &0 &0 &0 &  &  &  &  &  &  &  &  &  &  \\
K_{3}&-L_{3}&0 &0 &0 &0 &  &  &  &  &  &  &-K_{3}&K_{3}&  &  \\
  &   &-L_{3}&L_{3}&  &  &  &  &\dots&  &  &  &L_{3}&-L_{3}&  & \\
  &   &K_{3}&-K_{3}&  &  &  &  &  &  &0 &0 &0 &0 &-L_{3}&K_{3}\\
  &   &  &  &  &  &  &  &  &  &0 &0 &0 &0 &L_{3}&-K_{3} \\
  &   &  &  &  &  &  &  &0 &0 &0 &0 &0 &0 &0 &0 \\
  &   &  &  &  &  &  &  &0 &0 &0 &0 &0 &0 &0 &0 \\
  &   &  &  &  &  &-K_{3}&L_{3}&0 &0 &0 &0 &0 &0 &K_{3}&-L_{3}\\
  &   &  &  &  &  &K_{3}&-L_{3}&0 &0 &0 &0 &0 &0 &-K_{3}&L_{3}\\
  &   &  &  &  &  &  &  &-L_{3}&L_{3}&0 &0 &K_{3}&-K_{3}&0 &0 \\
  &   &  &  &  &  &  &  &K_{3}&-K_{3}&0 &0 &-L_{3}&L_{3}&0 &0 
\end{array}
\right),
\label{delta3}
\end{eqnarray}
\end{small}
where the matrix is composed of iterations of the matrices 
\\
$\displaystyle
\left(
\begin{array}{cc}
-K_{3}&K_{3}\\
L_{3}&-L_{3}
\end{array}
\right)
$, 
$\displaystyle
\left(
\begin{array}{cc}
-L_{3}&K_{3}\\
L_{3}&-K_{3}
\end{array}
\right)
$, 
and 
$\displaystyle
\left(
\begin{array}{cc}
-K_{3}&L_{3}\\
K_{3}&-L_{3}
\end{array}
\right)
$, 
$\displaystyle
\left(
\begin{array}{cc}
-L_{3}&L_{3}\\
K_{3}&-K_{3}
\end{array}
\right)
$. 

\newpage

The eigenvectors corresponding to the eigenvalue equal to $0$ are obtained as 
\vspace{1.2cm}
\begin{small}
\begin{eqnarray}
&
\left(
\begin{array}{c}
L_{1}\\
K_{1}\\
L_{2}\\
K_{2}\\
L_{3}\\
K_{3}\\
0\\
0\\
0\\
0\\
0\\
0\\
0\\
0\\
0\\
0\\
0\\
0\\
0\\
0\\
\vdots\\
0\\
0\\
0
\end{array}
\right),
\hspace{0.2cm}
\left(
\begin{array}{c}
0\\
0\\
0\\
\vdots\\
0\\
0\\
0\\
0\\
0\\
0\\
0\\
0\\
0\\
0\\
0\\
0\\
0\\
0\\
K_{3}\\
L_{3}\\
K_{2}\\
L_{2}\\
K_{1}\\
L_{1}
\end{array}
\right)
\hspace{1.4cm}
\left(
\begin{array}{c}
K_{1}+L_{3}\\
L_{1}+K_{3}\\
0\\
0\\
L_{1}\\
K_{1}\\
L_{2}\\
K_{2}\\
L_{3}\\
K_{3}\\
0\\
0\\
0\\
0\\
0\\
0\\
0\\
0\\
0\\
0\\
\vdots\\
0\\
0\\
0
\end{array}
\right),
\hspace{0.2cm}
\left(
\begin{array}{c}
K_{3}\\
L_{3}\\
K_{2}\\
L_{2}\\
K_{1}\\
L_{1}\\
0\\
0\\
L_{1}\\
K_{1}\\
L_{2}\\
K_{2}\\
L_{3}\\
K_{3}\\
0\\
0\\
0\\
0\\
0\\
0\\
\vdots\\
0\\
0\\
0
\end{array}
\right),
\hspace{0.2cm}
\left(
\begin{array}{c}
0\\
0\\
0\\
0\\

K_{3}\\
L_{3}\\
K_{2}\\
L_{2}\\
K_{1}\\
L_{1}\\
0\\
0\\
L_{1}\\
K_{1}\\
L_{2}\\
K_{2}\\
L_{3}\\
K_{3}\\
0\\
0\\
\vdots\\
0\\
0\\
0
\end{array}
\right),
\hspace{0.2cm}
\dots
\hspace{0.2cm},
\left(
\begin{array}{c}
0\\
0\\
0\\
\vdots\\
0\\
0\\
0\\
0\\
0\\
0\\
K_{3}\\
L_{3}\\
K_{2}\\
L_{2}\\
K_{1}\\
L_{1}\\
0\\
0\\
L_{1}\\
K_{1}\\
L_{2}\\
K_{2}\\
L_{3}\\
K_{3}
\end{array}
\right)
\hspace{0.2cm},
\left(
\begin{array}{c}
0\\
0\\
0\\
\vdots\\
0\\
0\\
0\\
0\\
0\\
0\\
0\\
0\\
0\\
0\\
K_{3}\\
L_{3}\\
K_{2}\\
L_{2}\\
K_{1}\\
L_{1}\\
0\\
0\\
L_{1}+K_{3}\\
K_{1}+L_{3}
\end{array}
\right)
\hspace{0.8cm}
\label{charges1}
\\
&
\hspace{0.4cm}
Q_{\rm Ia}
\hspace{1.2cm}
Q_{\rm Ib}
\hspace{2.8cm}
Q_{\rm II, 1}
\hspace{1.4cm}
Q_{\rm II, 2}
\hspace{1.2cm}
Q_{\rm II, 3}
\hspace{0.7cm}
\cdots
\hspace{0.7cm}
Q_{\rm II, \frac{M}{2}-2}
\hspace{1.0cm}
Q_{\rm II, \frac{M}{2}-1}
\hspace{0.8cm}
\nonumber
\end{eqnarray}
and 
\begin{eqnarray}
&
\left(
\begin{array}{c}
0\\
0\\
1\\
1\\
0\\
0\\
0\\
0\\
0\\
0\\
0\\
0\\
0\\
0\\
0\\
0\\
0\\
0\\
0\\
0\\
\vdots\\
0\\
0\\
0
\end{array}
\right),
\hspace{0.2cm}
\left(
\begin{array}{c}
0\\
0\\
0\\
0\\
0\\
0\\
1\\
1\\
0\\
0\\
0\\
0\\
0\\
0\\
0\\
0\\
0\\
0\\
0\\
0\\
\vdots\\
0\\
0\\
0
\end{array}
\right),
\hspace{0.2cm}
\dots
\hspace{0.2cm},
\left(
\begin{array}{c}
0\\
0\\
0\\
\vdots\\
0\\
0\\
0\\
0\\
0\\
0\\
0\\
0\\
0\\
0\\
0\\
0\\
0\\
0\\
0\\
0\\
1\\
1\\
0\\
0
\end{array}
\right)
\hspace{2.3cm}
\left(
\begin{array}{c}
1\\
0\\
0\\
0\\
0\\
1\\
0\\
0\\
1\\
0\\
0\\
0\\
0\\
1\\
0\\
0\\
1\\
0\\
0\\
0\\
0\\
1\\
\vdots\\
\vdots
\end{array}
\right),
\hspace{0.2cm}
\left(
\begin{array}{c}
0\\
1\\
0\\
0\\
1\\
0\\
0\\
0\\
0\\
1\\
0\\
0\\
1\\
0\\
0\\
0\\
0\\
1\\
0\\
0\\
1\\
0\\
\vdots\\
\vdots
\end{array}
\right).
\label{charges2}
\\
&
\hspace{1.6cm}
Q_{\rm III, 1}
\hspace{0.8cm}
Q_{\rm III, 2}
\hspace{0.5cm}
\cdots
\hspace{0.5cm}
Q_{\rm III, \frac{M}{2}-1}
\hspace{2.2cm}
Q_{\rm IVa}
\hspace{1.0cm}
Q_{\rm IVb}
\hspace{1.4cm}
\nonumber
\end{eqnarray}
\end{small}

Each of these eigenvectors represents a conserved charge of the Hamiltonian (\ref{HamGen}). 
The size of the matrix $\delta$ is $2(2N-1)$, 
and it is straightforward to convince 
that the rank of $\delta$ is $2(N-1)$. 
The eigenvalue $0$ is, therefore, $2(2N-1)-2(N-1)=2N$ fold degenerate, 
and this is the maximum number of the eigenvectors of $\delta$ 
corresponding to the eigenvalue $0$. 

There exist $M/2-1=N-1$ number of eigenvectors $Q_{\rm II}$, 
and also $N-1$ number of eigenvectors $Q_{\rm III}$, 
and together with $Q_{\rm Ia}$ and $Q_{\rm Ib}$, 
they usually form a complete basis set of $ {\rm Ker\:}\delta$, 
i.e. we have obtained all the conserved charges 
composed of the string type operators (\ref{base}).  

For some special parameters, however, 
$Q_{\rm Ia}$, $Q_{\rm Ib}$, and the series $Q_{\rm II}$, $Q_{\rm III}$ 
are not linearly independent. 
When we consider, for example, the case $K_{1}=L_{1}$ or $K_{1}=-L_{1}$, 
and $L_{2} =K_{2}=L_{3}=K_{3}=0$, 
they become linearly dependent, 
and we need $Q_{\rm IVa}$ or $Q_{\rm IVb}$  
to have a complete set of eigenvectors.

The charge $Q_{\rm Ia}$ is proportional to the Hamiltonian (\ref{HamGen}), 
and obviously commutes with (\ref{HamGen}). 
The series of conserved charges $Q_{\rm II}$ and $Q_{\rm III}$, 
in the case of $\eta^{(1)}_{j}$ and $L_{2}=K_{2}=L_{3}=K_{3}=0$,
i.e. in the case of the transverse Ising chain,  
were first derived in \cite{Grady82}. 
The charges $Q_{\rm II}$ and $Q_{\rm III}$,
in the case of $\eta^{(1)}_{j}$, 
with non-vanishing $L_{l}$ and $K_{l}$, 
were derived in \cite{Prosen98}. 
The transverse Ising chain and the XY chain 
on an infinite lattice ${\bf Z}$ were considered 
in \cite{AnshelevichGusev81} and in \cite{Gusev82}, respectively, 
and the conserved charges 
corresponding to $Q_{\rm II}$ and $Q_{\rm III}$ were derived. 
The charge $Q_{\rm Ib}$ does not appear in these arguments.

Let us consider the structures of the conserved charges. 
The conserved charges $Q_{\rm III}$ are written, 
for $m=1, 2, \ldots, N-1$, 
by the fermion operators $c_{1}(q)$ and $c_{2}(q)$, as 
\begin{eqnarray}
Q_{{\rm III}, m}
&=&
\sum_{j={\rm odd}}\eta_{j}\eta_{j+1}\cdots\eta_{j+2m-1}
+\sum_{j={\rm even}}\eta_{j}\eta_{j+1}\cdots\eta_{j+2m-1}
\hspace{0.6cm}
(m=1, 2, \ldots, N-1)
\nonumber
\\
&=&
2(-1)^{m}\sum_{j=1}^{N}\varphi_{1}(j)\varphi_{1}(j+m)
+2(-1)^{m}\sum_{j=1}^{N}\varphi_{2}(j)\varphi_{2}(j+m)
\nonumber
\\
&=&
2(-1)^{m}\sum_{0<q<\pi}
\Big(
e^{-imq}(c_{1}(q)c_{1}^{\dag}(q)+c_{2}(q)c_{2}^{\dag}(q))
+e^{imq}(c_{1}^{\dag}(q)c_{1}(q)+c_{2}^{\dag}(q)c_{2}(q))
\Big).
\end{eqnarray}
This can be expressed by the total number operator 
$c_{1}^{\dag}(q)c_{1}(q)+c_{2}^{\dag}(q)c_{2}(q)$, 
and therefore commute with the Hamiltonian ${\cal H}$, 
which can be written as a sum of 
hopping terms and the number operators of the fermions.

The conserved charge $Q_{\rm Ib}$ 
can be easily found from the symmetry 
of the representation matrices (\ref{delta1}), (\ref{delta2}) and (\ref{delta3}), 
but relates some more algebraic structures of the system.  
Uglov and Ivanov\cite{UglovIvanov96} 
introduced an ${\it sl}(n)$ analog of the Onsager algebra, 
and derived mutually commuting conserved quantities 
that satisfy a generalized Dolan-Grady condition. 
Their Hamiltonian is 
\begin{eqnarray}
{\cal H}_{\rm UI}
=
\sum_{j=1}^{n}k_j e_j
\hspace{0.6cm}(n\geq 3),
\label{hamUI}
\end{eqnarray}
where the operators $e_j$ satisfy the condition 
\begin{eqnarray}
[e_{i}, [e_{i}, e_{j}]]=e_{j}
\hspace{0.3cm}
(i\: {\rm and}\: j\: {\rm are\: adjacent}),
\hspace{0.6cm}
[e_{i}, e_{j}] =0
\hspace{0.3cm}
(i\: {\rm and}\: j\: {\rm are\: not\: adjacent}).
\label{condUI}
\end{eqnarray}
They introduced string operators 
\begin{eqnarray}
S_{j}(r)
=
[e_{j}, [e_{j+1}, [e_{j+2}, \cdots\hspace{0.2cm} [e_{j+r-2},\: e_{j+r-1}] \cdots ]\:]\:],
\end{eqnarray}
with $S_{j}(1)=e_{j}$ and $S_{j+n}(r)=S_{j}(r)$, 
and found conserved charges 
\begin{eqnarray}
I_{m}
=
\sum_{i=1}^{n}
k_{i}(S_{i}(nm+1)+2S_{i+1}(nm-1))
\hspace{0.3cm}
(m\geq 1), 
\hspace{0.6cm}
I_{0}={\cal H}_{UI}, 
\label{coschUI}
\end{eqnarray}
which satisfy 
$[I_{m}, I_{l}]=0$ $\:(m, l\geq 0)$. 

In our case, 
$\displaystyle e_{j}=\frac{1}{2}\eta_{j}$, 
with the periodic boundary condition, 
satisfy (\ref{condUI}). 
In this case, the string operators become 
\begin{eqnarray}
S_{j}(r)
=
\left\{
\begin{array}{ll}
\displaystyle \frac{1}{2}\eta_{j}\eta_{j+1}\cdots\eta_{j+r-1} & r<n \\
0    & r\geq n
\end{array}
\right.
\end{eqnarray}
In the case of $n=2N$, with the interactions 
$k_{i}=k_{\rm odd}$ $(i={\rm odd})$ and 
$k_{i}=k_{\rm even}$ $(i={\rm even})$, 
possible conserved charges are  
$I_{0}={\cal H}_{\rm UI}$ and $I_{1}$, 
where 
\begin{eqnarray}
I_{1}
=
k_{\rm odd}\sum_{j={\rm odd}}
\eta_{j}\eta_{j+1}\cdots\eta_{j+2N-2}
+
k_{\rm even}\sum_{j={\rm even}}
\eta_{j}\eta_{j+1}\cdots\eta_{j+2N-2}.
\end{eqnarray}
This $I_{1}$ is nothing but the conserved charge $Q_{\rm Ib}$, 
with $L_{2} =K_{2}=L_{3}=K_{3}=0$. 
Especially in the case 
$\displaystyle \eta^{(1)}_{2j-1}=\sigma^{z}_{j}$ and 
$\displaystyle \eta^{(1)}_{2j}=\sigma^{x}_{j}\sigma^{x}_{j+1}$, 
we find the transverse Ising chain, 
and in this case $I_{1}$ becomes 
\begin{eqnarray}
I_{1}
=
(-1)^{N-1}
\Big(
k_{\rm odd}\sum_{j=1}^{N}\prod_{l=j}^{j+N-2}\sigma^{z}_{l}
-k_{\rm even}\sum_{j=1}^{N}
\sigma^{y}_{j}(\prod_{l=j+1}^{j+N-2}\sigma^{z}_{l})\sigma^{y}_{j+N-1}
\Big).
\end{eqnarray}


The conserved charge $Q_{\rm Ib}$ 
can be understood in another way. 
From the relations 
$\displaystyle 
\eta_{j}\eta_{j+1}\cdots\eta_{j+2N-2}
=
(+2i)\varphi_{j-1}\varphi_{j+2N-2}$, 
$\displaystyle \varphi_{2j-2}=\varphi_{1}(j)$, 
$\displaystyle \varphi_{2j-1}=\varphi_{2}(j)$, 
and considering the Fourier transformation (\ref{Fourier}), 
it is easy to find that $Q_{\rm Ib}$ is obtained from $Q_{\rm Ia}$ 
multiplying $-1$ and replacing $q\mapsto -q$. 
Because of the relation 
\begin{eqnarray}
c_{\tau}(p)
&=&
\frac{1}{\sqrt{N}}\sum_{j=1}^{N}e^{-ipj}\varphi_{\tau}(j),
\nonumber
\\
c^{\dag}_{\tau}(p)
&=&
\frac{1}{\sqrt{N}}\sum_{j=1}^{N}e^{+ipj}\varphi_{\tau}(j)
=c_{\tau}(-p), 
\end{eqnarray}
it can be said 
that $Q_{\rm Ib}$ is obtained from $Q_{\rm Ia}$ 
by the particle-hole symmetry.

Let us again consider the operator $\eta_{1}\eta_{2}\cdots\eta_{M}=P$ 
in (\ref{bcond}),  
which commutes with the Hamiltonian ${\cal H}$, 
and with any products of $\eta_{j}$ $(j\geq 1)$. 
The total Hilbert space is divided into two subspaces 
in which $P=+1$ and $P=-1$, respectively. 
We find the relation 
$\eta_{j}\eta_{j+1}\cdots\eta_{j+l-1}
=(-1)^{l-1}P\eta_{j+l}\eta_{j+l+1}\cdots\eta_{M}\eta_{1}\cdots\eta_{j-1}$, 
and  
\begin{eqnarray}
\sum_{j={\rm even}}\eta_{j}\eta_{j+1}\cdots\eta_{j+(M-l)-1}
=
\left\{\begin{array}{ll}
\displaystyle 
P\sum_{j={\rm odd}}\eta_{j}\eta_{j+1}\cdots\eta_{j+l-1}&l={\rm odd}\\
\displaystyle 
(-1)P\sum_{j={\rm even}}\eta_{j}\eta_{j+1}\cdots\eta_{j+l-1}&l={\rm even}.\end{array}
\right.
\label{etaeqPeta}
\end{eqnarray}
This relation yields that 
the conserved charge $Q_{\rm IVa}$ can be written as 
\begin{eqnarray}
Q_{\rm IVa}
&=&
\sum_{j={\rm odd}}\eta_{j}
+\sum_{j={\rm even}}\eta_{j}\eta_{j+1}\eta_{j+2}
+\sum_{j={\rm odd}}\eta_{j}\eta_{j+1}\eta_{j+2}\eta_{j+3}\eta_{j+4}
+\cdots
\nonumber\\
&&
\hspace{3.4cm}
\cdots
+\sum_{j={\rm odd}}\eta_{j}\eta_{j+1}\cdots\eta_{j+M-4}
+\sum_{j={\rm even}}\eta_{j}\eta_{j+1}\cdots\eta_{j+M-2}
\nonumber
\\
&=&
(1+P)
\sum_{j={\rm odd}}
\Big(\:
\eta_{j}
+\eta_{j}\eta_{j+1}\eta_{j+2}\eta_{j+3}\eta_{j+4}
+\cdots
+\eta_{j}\eta_{j+1}\cdots\eta_{j+M-4}
\:\Big),
\end{eqnarray}
where we have used the fact that $N$ is assumed to be even. 
We then find that $Q_{\rm IVa}=0$ when $P=-1$. 
Similarly, we also find that $Q_{\rm IVb}=0$ with even $N$ and $P=-1$. 

Because of the relation (\ref{etaeqPeta}), 
nonlocal operators are sometimes transformed into local operators 
in each subspace with $P=+1$ or $P=-1$. 
For example $Q_{\rm Ib}$ can be rewritten as 
\begin{eqnarray}
Q_{\rm Ib}
&=&
L_{1}\sum_{j={\rm even}}\eta_{j}\eta_{j+1}\cdots\eta_{j+M-2}
+K_{1}\sum_{j={\rm odd}}\eta_{j}\eta_{j+1}\cdots\eta_{j+M-2}
\nonumber
\\
&+&
L_{2}\sum_{j={\rm even}}\eta_{j}\eta_{j+1}\cdots\eta_{j+M-3}
+K_{2}\sum_{j={\rm odd}}\eta_{j}\eta_{j+1}\cdots\eta_{j+M-3}
\nonumber
\\
&+&
L_{3}\sum_{j={\rm even}}\eta_{j}\eta_{j+1}\cdots\eta_{j+M-4}
+K_{3}\sum_{j={\rm odd}}\eta_{j}\eta_{j+1}\cdots\eta_{j+M-4}
\nonumber
\\
&=&
L_{1}P\sum_{j={\rm odd}}\eta_{j}
+K_{1}P\sum_{j={\rm even}}\eta_{j}
\nonumber
\\
&-&
L_{2}P\sum_{j={\rm even}}\eta_{j}\eta_{j+1}
-K_{2}P\sum_{j={\rm odd}}\eta_{j}\eta_{j+1}
\nonumber
\\
&+&
L_{3}P\sum_{j={\rm odd}}\eta_{j}\eta_{j+1}\eta_{j+2}
+K_{3}P\sum_{j={\rm even}}\eta_{j}\eta_{j+1}\eta_{j+2}
\nonumber
\\
&=&
PQ_{\rm Ia}-(L_2+K_2)PQ_{\rm III, 1}.
\end{eqnarray}
The operators $Q_{\rm Ia}$, $Q_{\rm Ib}$ and $Q_{\rm III, 1}$ are linearly independent 
in the subspace spanned by (\ref{base}), 
but relate with each other through $P$. 
Similarly, we can obtain the relations 
\begin{eqnarray}
PQ_{{\rm II}, 1}
&=&
Q_{{\rm II}, \frac{M}{2}-1} -(L_2+K_2)Q_{{\rm III}, \frac{M}{2}-3} ,
\nonumber
\\
PQ_{{\rm II}, m}
&=&
Q_{{\rm II}, \frac{M}{2}-m} 
-(L_2+K_2)(Q_{{\rm III}, m-1} +Q_{{\rm III}, m+1})
\hspace{0.9cm}
(m=2, \ldots, \frac{M}{2}-2),
\nonumber
\\
PQ_{{\rm III}, m}
&=&
-Q_{{\rm III}, \frac{M}{2}-m} 
\hspace{5.8cm}
(m=1, 2, \ldots, \frac{M}{2}-1).
\end{eqnarray}

It should be noted that 
because of the relation (\ref{etaeqPeta}), 
and because of the relation 
between $\eta_{j}$ and physical operators 
such as the Pauli operators, 
operators (\ref{base}) are sometimes not independent. 
In this case, 
some of the charges derived as the eigenvectors of $\delta$ 
become dependent, or become $0$.

All the charges are written, through the Fourier transform, 
by the fermi operators 
$c_{1}^{\dag}(q)c_{1}(q)$, $c_{2}^{\dag}(q)c_{2}(q)$, 
$c_{1}^{\dag}(q)c_{2}(q)$, and $c_{1}(q)c_{2}^{\dag}(q)$. 
From this fact, and the symmetry of the system, 
it is straightforward to convince 
that all the charges commute with each other. 

Let us write down some examples of 
Hamiltonians and corresponding conserved charges explicitly.  
In the case of the series $\{\eta^{(1)}_{j}\}$, 
i.e. the series $(1, 0, 1)$ in Table 1 and Table 2, 
which contains the transverse Ising chain, 
and the XY and the cluster interactions, 
we find the solvable Hamiltonian 
\begin{eqnarray}
-\beta{\cal H}_{\rm TI}
&=&
\underline{
L_{1}\sum_{j=1}^{N} \sigma^{z}_{j}
+K_{1}\sum_{j=1}^{N} \sigma^{x}_{j}\sigma^{x}_{j+1}
}
+L_{2}\sum_{j=1}^{N} (+i)\:\sigma^{y}_{j}\sigma^{x}_{j+1}
+K_{2}\sum_{j=1}^{N} (-i)\:\sigma^{x}_{j}\sigma^{y}_{j+1}
\nonumber
\\
&+&
\underline{
L_{3}\sum_{j=1}^{N} \sigma^{y}_{j}\sigma^{y}_{j+1}
+
K_{3}\sum_{j=1}^{N} (-1)\:\sigma^{x}_{j}\sigma^{z}_{j+1}\sigma^{x}_{j+2}
}.
\end{eqnarray}
Corresponding to this Hamiltonian, 
we find, for example, 
the first charge of the series $Q_{\rm II}$ represented by $Q_{\rm II, 1}$ is 
\begin{eqnarray}
Q_{\rm II, 1}
&=&
(K_{1}+L_{3})\sum_{j=1}^{N} \sigma^{z}_{j}
+(L_{1}+K_{3})\sum_{j=1}^{N} \sigma^{x}_{j}\sigma^{x}_{j+1}
+L_{1}\sum_{j=1}^{N} \sigma^{y}_{j}\sigma^{y}_{j+1}
+K_{1}\sum_{j=1}^{N} (-1)\:\sigma^{x}_{j}\sigma^{z}_{j+1}\sigma^{x}_{j+2}
\nonumber
\\
&&
+L_{2}\sum_{j=1}^{N} (-i)\:\sigma^{y}_{j}\sigma^{z}_{j+1}\sigma^{x}_{j+2}
+K_{2}\sum_{j=1}^{N} (+i)\:\sigma^{x}_{j}\sigma^{z}_{j+1}\sigma^{y}_{j+2}
\nonumber
\\
&&
+L_{3}\sum_{j=1}^{N} (-1)\:\sigma^{y}_{j}\sigma^{z}_{j+1}\sigma^{y}_{j+2}
+K_{3}\sum_{j=1}^{N} \sigma^{x}_{j}(\sigma^{z}_{j+1}\sigma^{z}_{j+2})\sigma^{x}_{j+3},
\end{eqnarray}
and $[-\beta{\cal H}_{\rm TI}, Q_{\rm II, 1}]=0$ is satified. 

In the case of the series $\{\eta^{(2)}_{j}\}$, and $\{\zeta^{(2)}_{j}\}$, 
i.e. the series $(2, 1, 0)$ in Table 1 and Table 2, 
we can introduce the XY chain as shown in (\ref{hamXY}). 
From (\ref{base}) and (\ref{charges1})-(\ref{charges2}), 
we can derive a conserved charge $Q^{\eta}$ 
corresponding to ${\cal H^{\eta}}$ in (\ref{hamXY}), 
where $[{\cal H^{\eta}}, Q^{\eta}]=0$ is satisfied. 
This $Q^{\eta}$ also satisfies $[{\cal H^{\zeta}}, Q^{\eta}]=0$
because the operators 
$\eta_{j}^{(2)}$ $(j\geq 1)$ and $\zeta_{k}^{(2)}$ $(k\geq 1)$ 
commute with each other, 
and as a result, 
$[{\cal H}, Q^{\eta}]=0$, 
${\cal H}={\cal H^{\eta}}+{\cal H^{\zeta}}$. 
From $\{\eta^{(2)}_{j}\}$, and $\{\zeta^{(2)}_{j}\}$, 
we can introduce a solvable Hamiltonian, for example  
\begin{eqnarray}
-\beta{\cal H}_{\rm XY}
&=&
\underline{
L_{1}\sum_{j=1}^{N} \sigma^{x}_{j}\sigma^{x}_{j+1}
+K_{1}\sum_{j=1}^{N} \sigma^{y}_{j}\sigma^{y}_{j+1}
}
\nonumber\\
&&
+L_{2}\sum_{j=1}^{N} (+i)\:\sigma^{x}_{j}\sigma^{z}_{j+1}\sigma^{y}_{j+2}
+K_{2}\sum_{j=1}^{N} (-i)\:\sigma^{y}_{j}\sigma^{z}_{j+1}\sigma^{x}_{j+2}
\nonumber\\
&&
+L_{3}\sum_{j=1}^{N} \sigma^{x}_{j}(\sigma^{z}_{j+1}\sigma^{z}_{j+2})\sigma^{x}_{j+3}
+K_{3}\sum_{j=1}^{N} \sigma^{y}_{j}(\sigma^{z}_{j+1}\sigma^{z}_{j+2})\sigma^{y}_{j+3}.
\end{eqnarray}
Corresponding to this Hamiltonian, 
we find, for example,  
\begin{eqnarray}
Q_{\rm II, 1}
&=&
(K_{1}+L_{3})\sum_{j=1}^{N} \sigma^{x}_{j}\sigma^{x}_{j+1}
+(L_{1}+K_{3})\sum_{j=1}^{N} \sigma^{y}_{j}\sigma^{y}_{j+1}
\nonumber
\\
&&
+L_{1}\sum_{j=1}^{N} \sigma^{x}_{j}(\sigma^{z}_{j+1}\sigma^{z}_{j+2})\sigma^{x}_{j+3}
+K_{1}\sum_{j=1}^{N} \sigma^{y}_{j}(\sigma^{z}_{j+1}\sigma^{z}_{j+2})\sigma^{y}_{j+3}
\nonumber
\\
&&
+L_{2}\sum_{j=1}^{N} (+i)\:
\sigma^{x}_{j}(\sigma^{z}_{j+1}\sigma^{z}_{j+2}\sigma^{z}_{j+3})\sigma^{y}_{j+4}
+K_{2}\sum_{j=1}^{N} (-i)\:
\sigma^{y}_{j}(\sigma^{z}_{j+1}\sigma^{z}_{j+2}\sigma^{z}_{j+3})\sigma^{x}_{j+4}
\nonumber
\\
&&
+L_{3}\sum_{j=1}^{N} 
\sigma^{x}_{j}(\sigma^{z}_{j+1}\sigma^{z}_{j+2}\sigma^{z}_{j+3}\sigma^{z}_{j+4})\sigma^{x}_{j+5}
+K_{3}\sum_{j=1}^{N} 
\sigma^{y}_{j}(\sigma^{z}_{j+1}\sigma^{z}_{j+2}\sigma^{z}_{j+3}\sigma^{z}_{j+4})\sigma^{y}_{j+5}.
\end{eqnarray}
In the case of the series $\{\eta^{(3)}_{j}\}$, 
i.e. the series $(3, 1, 1)$ in Table 1 and Table 2, 
which contain the cluster interactions 
and the next-nearest-neighbor $x$-$x$ interactions, 
we find a solvable Hamiltonian  
\begin{eqnarray}
-\beta{\cal H}_{\rm cl,x}
&=&
\underline{
L_{1}\sum_{j=1}^{N} \sigma^{x}_{j}\sigma^{z}_{j+1}\sigma^{x}_{j+2}
+K_{1}\sum_{j=1}^{N} \sigma^{x}_{j}1_{j+1}\sigma^{x}_{j+2}
}
\nonumber
\\
&&
+L_{2}\sum_{j=1}^{N} (+i)\:\sigma^{x}_{j}\sigma^{y}_{j+1}\sigma^{x}_{j+2}\sigma^{x}_{j+3}
+K_{2}\sum_{j=1}^{N} (-i)\:\sigma^{x}_{j}\sigma^{x}_{j+1}\sigma^{y}_{j+2}\sigma^{x}_{j+3}
\nonumber
\\
&&
+L_{3}\sum_{j=1}^{N} 
\sigma^{x}_{j}\sigma^{y}_{j+1}1_{j+2}\sigma^{y}_{j+3}\sigma^{x}_{j+4}
+K_{3}\sum_{j=1}^{N} 
(-1)\:\sigma^{x}_{j}\sigma^{x}_{j+1}\sigma^{z}_{j+2}\sigma^{x}_{j+3}\sigma^{x}_{j+4}.
\end{eqnarray}
Corresponding to this Hamiltonian, 
we find, for example,  
\begin{eqnarray}
Q_{\rm II, 1}
&=&
(K_{1}+L_{3})\sum_{j=1}^{N} \sigma^{x}_{j}\sigma^{z}_{j+1}\sigma^{x}_{j+2}
+(L_{1}+K_{3})\sum_{j=1}^{N} \sigma^{x}_{j}1_{j+1}\sigma^{x}_{j+2}
\nonumber
\\
&&
+L_{1}\sum_{j=1}^{N} 
\sigma^{x}_{j}\sigma^{y}_{j+1}1_{j+2}\sigma^{y}_{j+3}\sigma^{x}_{j+4}
+K_{1}\sum_{j=1}^{N}
(-1)\:\sigma^{x}_{j}\sigma^{x}_{j+1}\sigma^{z}_{j+2}\sigma^{x}_{j+3}\sigma^{x}_{j+4}
\nonumber
\\
&&
+L_{2}\sum_{j=1}^{N} 
(-i)\:\sigma^{x}_{j}\sigma^{y}_{j+1}(1_{j+2}\sigma^{z}_{j+3})\sigma^{x}_{j+4}\sigma^{x}_{j+5}
+K_{2}\sum_{j=1}^{N} 
(+i)\:\sigma^{x}_{j}\sigma^{x}_{j+1}\sigma^{z}_{j+2}1_{j+3}\sigma^{y}_{j+4}\sigma^{x}_{j+5}
\nonumber
\\
&&
+L_{3}\sum_{j=1}^{N} 
(-1)\:\sigma^{x}_{j}\sigma^{y}_{j+1}(1_{j+2}\sigma^{z}_{j+3})1_{j+4}\sigma^{y}_{j+5}\sigma^{x}_{j+6}
+K_{3}\sum_{j=1}^{N} 
\sigma^{x}_{j}\sigma^{x}_{j+1}\sigma^{z}_{j+2}(1_{j+3}\sigma^{z}_{j+4})\sigma^{x}_{j+5}\sigma^{x}_{j+6}.
\nonumber\\
\end{eqnarray}

In the case of the series $\{\eta^{(4)}_{j, k}\}$, 
i.e. the series $(4, 1, -)$ in Table 1 and Table 2, 
which contain the cluster interactions 
and the next-nearest-neighbor $z$-$z$ interactions, 
we find a solvable Hamiltonian  
\begin{eqnarray}
-\beta{\cal H}_{\rm cl,z}
&=&
\underline{
L_{1}\sum_{j=1}^{N} \sigma^{x}_{j}\sigma^{z}_{j+1}\sigma^{x}_{j+2}
+K_{1}\sum_{j=1}^{N} \sigma^{z}_{j}1_{j+1}\sigma^{z}_{j+2}
\nonumber
}\\
&&
+L_{2}\sum_{j=1}^{N} (-i)\:
\sigma^{x}_{j}\sigma^{z}_{j+1}\sigma^{y}_{j+2}1_{j+3}\sigma^{z}_{j+4}
+K_{2}\sum_{j=1}^{N} (+i)\:
\sigma^{z}_{j}1_{j+1}\sigma^{y}_{j+2}\sigma^{z}_{j+3}\sigma^{x}_{j+4}
\nonumber
\\
&&
+L_{3}\sum_{j=1}^{N} 
\sigma^{x}_{j}\sigma^{z}_{j+1}\sigma^{y}_{j+2}1_{j+3}\sigma^{y}_{j+4}\sigma^{z}_{j+5}\sigma^{x}_{j+6}
+K_{3}\sum_{j=1}^{N} 
\sigma^{z}_{j}1_{j+1}(\sigma^{y}_{j+2}\sigma^{z}_{j+3}\sigma^{y}_{j+4}1_{j+5})\sigma^{z}_{j+6}.
\nonumber
\\
\end{eqnarray}
Corresponding to this Hamiltonian, 
we find, for example,  
\begin{eqnarray}
Q_{\rm II, 1}
&=&
(K_{1}+L_{3})\sum_{j=1}^{N} \sigma^{x}_{j}\sigma^{z}_{j+1}\sigma^{x}_{j+2}
+(L_{1}+K_{3})\sum_{j=1}^{N} \sigma^{z}_{j}1_{j+1}\sigma^{z}_{j+2}
\nonumber
\\
&&
+L_{1}\sum_{j=1}^{N} 
\sigma^{x}_{j}\sigma^{z}_{j+1}\sigma^{y}_{j+2}1_{j+3}\sigma^{y}_{j+4}\sigma^{z}_{j+5}\sigma^{x}_{j+6}
+K_{1}\sum_{j=1}^{N}
\sigma^{z}_{j}1_{j+1}(\sigma^{y}_{j+2}\sigma^{z}_{j+3}\sigma^{y}_{j+4}1_{j+5})\sigma^{z}_{j+6}\nonumber
\\
&&
+L_{2}\sum_{j=1}^{N} (-i)\:
\sigma^{x}_{j}\sigma^{z}_{j+1}\sigma^{y}_{j+2}1_{j+3}(\sigma^{y}_{j+4}\sigma^{z}_{j+5}\sigma^{y}_{j+6}1_{j+7})\sigma^{z}_{j+8}
\nonumber
\\
&&
+K_{2}\sum_{j=1}^{N} (+i)\:
\sigma^{z}_{j}1_{j+1}(\sigma^{y}_{j+2}\sigma^{z}_{j+3}\sigma^{y}_{j+4}1_{j+5})\sigma^{y}_{j+6}\sigma^{z}_{j+7}\sigma^{x}_{j+8}
\nonumber
\\
&&
+L_{3}\sum_{j=1}^{N} 
\sigma^{x}_{j}\sigma^{z}_{j+1}\sigma^{y}_{j+2}1_{j+3}(\sigma^{y}_{j+4}\sigma^{z}_{j+5}\sigma^{y}_{j+6}1_{j+7})\sigma^{y}_{j+8}\sigma^{z}_{j+9}\sigma^{x}_{j+10}
\nonumber
\\
&&
+K_{3}\sum_{j=1}^{N} 
\sigma^{z}_{j}1_{j+1}(\sigma^{y}_{j+2}\sigma^{z}_{j+3}\sigma^{y}_{j+4}1_{j+5})(\sigma^{y}_{j+6}\sigma^{z}_{j+7}\sigma^{y}_{j+8}1_{j+9})\sigma^{z}_{j+10}.
\end{eqnarray}

One can find, for example in Table 1 and Table 2, 
an infinite number of examples of Hamiltonians 
that satisfy the condition, 
and correspondingly one can write down the conserved charges 
corresponding to each Hamiltonian.

\section{Conclusions}

We consider solvable Hamiltonian (\ref{HamGen}) 
that satisfy the condition (\ref{cond}). 
This system can be mapped to the free fermion system. 
We found all the conserved charges of (\ref{HamGen}) 
written by the string-type products of the interactions, 
i.e. two trivial charges $1$ and $\eta_{1}\eta_{2}\cdots\eta_{M}$, 
and $2N$ charges 
which are obtained as eigenvectors with eigenvalue $0$ 
of the representation matrix $\delta$. 
One can use the standard technique of linear algebra, 
and one can also determine 
the dimension of ${\rm Ker\:} \delta$ 
from the rank of the matrix $\delta$, 
and hence can determine the maximum number of conserved charges, 
and be able to show 
that all the conserved charges in this subspace are obtained. 
In the case of the transverse Ising chain, 
all the known charges are re-derived, 
and in the case of other models that satisfy (\ref{cond}), 
new conserved charges are derived. 
\\

\section*{Acknowledgements}
This work was supported by JSPS KAKENHI Grant Number 19K03668.
\\



\begin{table}
\caption{\label{table1}
Examples of solvable Hamiltonians 
$-\beta{\cal H}(k, n, l)
=L_{1}\sum_{j=1}^{N}\eta_{2j-1}+K_{1}\sum_{j=1}^{N}\eta_{2j}$, 
where $n$ and $l$ are arbitrary integers. 
(From Table 1 in \cite{Minami17}.) 
}
\footnotesize
\begin{tabular}{@{}lll}
\hline
\hline
$(k, n, l)$ & 
$-\beta{\cal H}(k, n, l)$
&
\\
\hline
$(1, n, l)$ & 
$\displaystyle 
L_1\sum_{j=1}^N
\Big(\prod_{\nu=1}^{n}\sigma_{j+\nu-1}^x\Big)
\Big(\prod_{\nu=1}^{l}\sigma_{j+n+\nu-1}^z\Big)
\Big(\prod_{\nu=1}^{n}\sigma_{j+n+l+\nu-1}^x\Big)
$
&
$\displaystyle 
+
K_1\sum_{j=1}^N
\sigma_{j}^x
\sigma_{j+1}^x
$ \\
$(2, n, l)$ & 
$\displaystyle 
L_1\sum_{j=1}^N
\Big(\prod_{\nu=1}^{n}\sigma_{j+\nu-1}^x\Big)
\Big(\prod_{\nu=1}^{l}\sigma_{j+n+\nu-1}^z\Big)
\Big(\prod_{\nu=1}^{n}\sigma_{j+n+l+\nu-1}^x\Big)
$
&
$\displaystyle 
+
K_1\sum_{j=1}^N
\sigma_{j}^y
\sigma_{j+1}^y
$ \\
$(3, n, l)$ & 
$\displaystyle 
L_1\sum_{j=1}^N
\Big(\prod_{\nu=1}^{n}\sigma_{j+\nu-1}^x\Big)
\sigma_{j+n}^z
\Big(\prod_{\nu=1}^{n}\sigma_{j+n+\nu}^x\Big)
$
&
$\displaystyle 
+
K_1\sum_{j=1}^N
\sigma_{j}^x
\Big(\prod_{\nu=1}^{l}{\bf 1}_{j+\nu}\Big)
\sigma_{j+l+1}^x
$ \\
$(4, n, -)$ & 
$\displaystyle 
L_1\sum_{j=1}^N
\sigma_{j}^x
\Big(\prod_{\nu=1}^{n}\sigma_{j+\nu}^z\Big)
\sigma_{j+n+1}^x
$
&
$\displaystyle 
+
K_1\sum_{j=1}^N
\sigma_{j}^z
\Big(\prod_{\nu=1}^{n}{\bf 1}_{j+\nu}\Big)
\sigma_{j+n+1}^z
$ \\
$(5, n, -)$ & 
$\displaystyle 
L_1\sum_{j=1}^N
\Big(\prod_{\nu=1}^{n}\sigma_{j+\nu-1}^z\Big)
$
&
$\displaystyle 
+
K_1\sum_{j=1}^N
\sigma_{j}^x
\sigma_{j+1}^x
$ \\
$(6, n, -)$ & 
$\displaystyle 
L_1\sum_{j=1}^N
\sigma_{j}^x
\Big(\prod_{\nu=1}^{n}{\bf 1}_{j+\nu}\Big)
\sigma_{j+n+1}^x
$
&
$\displaystyle 
+
K_1\sum_{j=1}^N
\sigma_{j}^z
$ \\
$(7, n, l)$ & 
$\displaystyle 
L_1\sum_{j=1}^N
\sigma_{j}^x
\Big(\prod_{\nu=1}^{n}\sigma_{j+\nu}^z\Big)
\sigma_{j+n+1}^x
$
&
$\displaystyle 
+
K_1\sum_{j=1}^N
\sigma_{j}^x
\Big(\prod_{\nu=1}^{l}\sigma_{j+\nu}^z\Big)
\sigma_{j+l+1}^x
$ \\
$(8, n, l)$ & 
$\displaystyle 
L_1\sum_{j=1}^N
\sigma_{j}^x
\Big(\prod_{\nu=1}^{n}\sigma_{j+\nu}^z\Big)
\sigma_{j+n+1}^x
$
&
$\displaystyle 
+
K_1\sum_{j=1}^N
\sigma_{j}^y
\Big(\prod_{\nu=1}^{l}\sigma_{j+\nu}^z\Big)
\sigma_{j+l+1}^y
$ \\
$(9, n, -)$ & 
$\displaystyle 
L_1\sum_{j=1}^N
\sigma_{j}^x\sigma_{j+1}^x\sigma_{j+2}^x
\Big(\prod_{\nu=1}^{n}\sigma_{j+2+\nu}^z\Big)
\sigma_{j+n+3}^x\sigma_{j+n+4}^x\sigma_{j+n+5}^x
$
&
$\displaystyle 
+
K_1\sum_{j=1}^N
\sigma_{j}^x
\Big(\prod_{\nu=1}^{n+2}\sigma_{j+\nu}^z\Big)
\sigma_{j+n+3}^x
$ \\
$(10, n, -)$ & 
$\displaystyle 
L_1\sum_{j=1}^N
\sigma_{j}^x\sigma_{j+1}^x
\Big(\prod_{\nu=1}^{n}\sigma_{j+1+\nu}^z\Big)
\sigma_{j+n+2}^x\sigma_{j+n+3}^x
$
&
$\displaystyle 
+
K_1\sum_{j=1}^N
\Big(\prod_{\nu=1}^{n+2}\sigma_{j+\nu-1}^z\Big)
$ \\
$(11, -, -)$ & 
$\displaystyle 
L_1\sum_{j=1}^N
\sigma_{j}^x\sigma_{j+1}^x
\sigma_{j+2}^z
\sigma_{j+3}^x\sigma_{j+4}^x
$
&
$\displaystyle 
+
K_1\sum_{j=1}^N
\sigma_{j}^x
\sigma_{j+1}^z
\sigma_{j+2}^x
$ \\
\hline
\hline
\end{tabular}\\

\end{table}


\begin{table}
\caption{\label{table2}
Every $(k, n, l)$ provides a series solvable interactions, 
where $(k, n, l)$ is the index given in Table 1. 
Six interactions 
which are obtained from string-type products of $\eta_{j}$ 
are shown explicitly.  
The initial operator $\eta^{(k, n, l)}_0$ 
and the transformation $\varphi^{(k, n, l)}_{j}$ 
which diagonalize the Hamiltonian are shown in the last row. 
The first case $(k, n, l)=(1, 0, 1)$ 
includes the transverse Ising model, the XY model, and the cluster model, as special cases. 
The second case $(k, n, l)=(2, 1, 0)$ 
includes the one-dimensional Kitaev model. 
The last two cases $(3, 1, 1)$ and $(4, 1, -)$ 
cannot be solved by the Jordan-Wigner transformation. 
(See also Table 2 in \cite{Minami17}.)
}
\footnotesize
\begin{tabular}{@{}ll}
&
\\
\multicolumn{2}{l}{$(k, n, l)$}
\\
\hline
\hline
$-2i\varphi^{(k, n, l)}_2(j)\varphi^{(k, n, l)}_1(j-1)
\hspace{0.1cm}
=\eta^{(k, n, l)}_{2j-3}\eta^{(k, n, l)}_{2j-2}\eta^{(k, n, l)}_{2j-1}$ & 
\\
$+2i\varphi^{(k, n, l)}_2(j)\varphi^{(k, n, l)}_1(j)
\hspace{0.55cm}
=\eta^{(k, n, l)}_{2j-1}$ & 
\hspace{0.2cm}
$+2i\varphi^{(k, n, l)}_1(j)\varphi^{(k, n, l)}_1(j+1)
=-i\:\eta^{(k, n, l)}_{2j-1}\eta^{(k, n, l)}_{2j}$ 
\\
$-2i\varphi^{(k, n, l)}_2(j)\varphi^{(k, n, l)}_1(j+1) 
\hspace{0.1cm}
=\eta^{(k, n, l)}_{2j}$ & 
\hspace{0.2cm}
$-2i\varphi^{(k, n, l)}_2(j)\varphi^{(k, n, l)}_2(j+1)
=i\:\eta^{(k, n, l)}_{2j}\eta^{(k, n, l)}_{2j+1}$ 
\\
$+2i\varphi^{(k, n, l)}_2(j)\varphi^{(k, n, l)}_1(j+2)
\hspace{0.1cm}
=\eta^{(k, n, l)}_{2j}\eta^{(k, n, l)}_{2j+1}\eta^{(k, n, l)}_{2j+2}$ & 
\\
\hline
$\eta^{(k, n, l)}_0$ &
$\varphi^{(k, n, l)}_{2j}$ 
\hspace{0.4cm}
and 
\hspace{0.6cm}
$\varphi^{(k, n, l)}_{2j+1}$ 
\hspace{0.8cm}
$(j=0, 1, 2, 3, \ldots)$
\hspace{0.6cm}
\\
\hline
\end{tabular}

\begin{tabular}{@{}ll}
&
\\
&
\\
\multicolumn{2}{l}{$(1, 0, 1)$}
\\
\hline
$\sigma_{j-1}^y\sigma_{j}^y$ & 
\\
$\sigma^z_{j}$ & 
\hspace{1.2cm}
$\sigma_{j}^y\sigma_{j+1}^x$ 
\\
$\sigma_{j}^x\sigma_{j+1}^x$ & 
\hspace{1.2cm}
$\sigma_{j}^x\sigma_{j+1}^y$ 
\\
$(-1)\sigma_{j}^x\sigma_{j+1}^z\sigma_{j+2}^x$ & 
\\
\hline
$i\sigma^x_{1}$ &
$\displaystyle \varphi^{(1, 0, 1)}_{2j}=\frac{1}{\sqrt{2}}(\prod_{\nu=1}^{j}\sigma^z_{\nu})\sigma^x_{j+1}$ 
\hspace{0.6cm}
$\displaystyle \varphi^{(1, 0, 1)}_{2j+1}=\frac{1}{\sqrt{2}}(\prod_{\nu=1}^{j}\sigma^z_{\nu})\sigma^y_{j+1}$
\\
\hline
&
\\
\multicolumn{2}{l}{$(2, 1, 0)$}
\\
\hline
$\sigma_{2j-3}^x\sigma_{2j-2}^z\sigma_{2j-1}^z\sigma_{2j}^x$ & 
\\
$\sigma^x_{2j-1}\sigma^x_{2j}$ & 
\hspace{1.2cm}
$\sigma_{2j-1}^x\sigma_{2j}^z\sigma_{2j+1}^y$ 
\\
$\sigma^y_{2j}\sigma^y_{2j+1}$ & 
\hspace{1.2cm}
$\sigma_{2j}^y\sigma_{2j+1}^z\sigma_{2j+2}^x$ 
\\
$\sigma_{2j}^y\sigma_{2j+1}^z\sigma_{2j+2}^z\sigma_{2j+3}^y$ & 
\\
\hline
$i\sigma^y_{1}$ &
$\displaystyle \varphi^{(2, 1, 0)}_{2j}=\frac{(-1)^j}{\sqrt{2}}(\prod_{\nu=1}^{2j}\sigma^z_{\nu})\sigma^y_{2j+1}$ 
\hspace{0.6cm}
$\displaystyle \varphi^{(2, 1, 0)}_{2j+1}=\frac{(-1)^j}{\sqrt{2}}(\prod_{\nu=1}^{2j+1}\sigma^z_{\nu})\sigma^x_{2j+2}$
\\
\hline
\end{tabular}
\end{table}

\begin{table}
\footnotesize
\begin{tabular}{@{}ll}
&
\\
\multicolumn{2}{l}{$(3, 1, 1)$}
\\
\hline
$\sigma_{2j-3}^x\sigma_{2j-2}^y 1_{2j-1}\sigma_{2j}^y\sigma_{2j+1}^x$ & 
\\
$\sigma_{2j-1}^x\sigma_{2j}^z\sigma_{2j+1}^x$ & 
$\sigma_{2j-1}^x\sigma_{2j}^y\sigma_{2j+1}^x\sigma_{2j+2}^x$ 
\\
$\sigma_{2j}^x 1_{2j+1}\sigma_{2j+2}^x$ & 
$\sigma_{2j}^x\sigma_{2j+1}^x\sigma_{2j+2}^y\sigma_{2j+3}^x$ 
\\
$(-1)\sigma_{2j}^x\sigma_{2j+1}^x\sigma_{2j+2}^z\sigma_{2j+3}^x\sigma_{2j+4}^x$ & 
\\
\hline
$i\sigma^x_{2}$ &
$\displaystyle \varphi^{(3, 1, 1)}_{2j}
=\frac{1}{\sqrt{2}}\sigma^x_{1}(\prod_{\nu=1}^{j}1_{2\nu-1}\sigma^z_{2\nu})\sigma^x_{2j+1}\sigma^x_{2j+2}$ 
\\
 &
$\displaystyle \varphi^{(3, 1, 1)}_{2j+1}
=\frac{1}{\sqrt{2}}\sigma^x_{1}(\prod_{\nu=1}^{j}1_{2\nu-1}\sigma^z_{2\nu})1_{2j+1}\sigma^y_{2j+2}\sigma^x_{2j+3}$
\\

&
\\
\multicolumn{2}{l}{$(4, 1, -)$}
\\
\hline
$\sigma_{4j-7}^x\sigma_{4j-6}^z \sigma_{4j-5}^y 1_{4j-4}\sigma_{4j-3}^y\sigma_{4j-2}^z\sigma_{4j-1}^x$ & 
\\
$\sigma_{4j-3}^x\sigma_{4j-2}^z\sigma_{4j-1}^x$ & 
$(-1)\sigma_{4j-3}^x\sigma_{4j-2}^z\sigma_{4j-1}^y 1_{4j}\sigma_{4j+1}^z$ 
\\
$\sigma_{4j-1}^z 1_{4j}\sigma_{4j+1}^z$ & 
$(-1)\sigma_{4j-1}^z 1_{4j}\sigma_{4j+1}^y\sigma_{4j+2}^z\sigma_{4j+3}^x$ 
\\
$\sigma_{4j-1}^z 1_{4j}\sigma_{4j+1}^y\sigma_{4j+2}^z\sigma_{4j+3}^y 1_{4j+4}\sigma_{4j+5}^z$ & 
\\
\hline
$i\sigma^x_{2}$ 
&
$\displaystyle \varphi^{(4, 1, -)}_{2j}
=\frac{(-1)^{j}}{\sqrt{2}}\sigma^z_{1}\sigma^x_{2}
(\prod_{\nu=1}^{j}\sigma^y_{4\nu-3}\sigma^z_{4\nu-2}\sigma^y_{4\nu-1}1_{4\nu})\sigma^z_{4j+1}$ 
\\
&
$\displaystyle \varphi^{(4, 1, -)}_{2j+1}
=\frac{(-1)^{j+1}}{\sqrt{2}}\sigma^z_{1}\sigma^x_{2}
(\prod_{\nu=1}^{j}\sigma^y_{4\nu-3}\sigma^z_{4\nu-2}\sigma^y_{4\nu-1}1_{4\nu})\sigma^y_{4j+1}\sigma^z_{4j+2}\sigma^x_{4j+3}$ 
\\
\hline
\end{tabular}
\end{table}

\end{document}